\documentclass[11pt]{article}
\usepackage{amsmath}
\usepackage{amssymb}
\usepackage{graphicx}
\usepackage[dvips]{color}
\usepackage{dcolumn}
\usepackage{bm}
\input mysym
\begin{document}
\begin{figure}
\leftline{\includegraphics[scale=0.5]{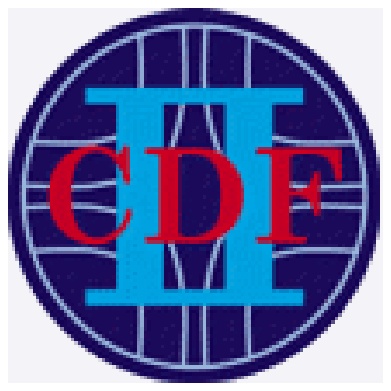}\hfill
FERMILAB-CONF-04-424-E }
\end{figure}

\title{Measurement of Orbitally Excited D-Mesons \\
       at CDF II}
\author{Igor V. Gorelov \thanks{talk given on behalf of the CDF
 Collaboration at the First Meeting of the APS Topical Group on Hadronic 
 Physics,~GHP~2004,~Oct~24-26,~2004,~Fermilab,~Batavia,~Illinois.}\\
       (For the CDF Collaboration) \\
 \small{\textit{Department of Physics and Astronomy,}}\\ 
 \small{\textit{University of New Mexico,}}\\ 
 \small{\textit{800 Yale Blvd. NE, Albuquerque, NM 87131, USA}}\\
 \small{\textit{email:gorelov@fnal.gov}}  
}
\date{}
\maketitle
\begin{abstract}
  Results of the first measurement of $^{3}P$ orbitally excited neutral
  $D$-meson states, \Dorbtwoz and \Dorbonez, produced in hadron
  collisions at the Tevatron are presented.  Using data from the displaced
  track trigger, CDF II collects a large sample of these
  states in decay modes \Dstarp\pim, \Dp\pim. Masses and widths of
  both states have been measured with precision better than or
  comparable to that of the world average.
\end{abstract}.
\section{Introduction}
  The mesons containing one heavy quark are a useful laboratory to
  test QCD models.  In the limit of heavy quark mass
  $m_{Q}\to\infty$, heavy mesons' properties are governed by the dynamics
  of the light quark. As such, these states become ``hydrogen
  atoms'' of hadron physics. In this Heavy Quark Symmetry approach 
  (see References~\cite{th:isgur1,th:rosner,th:isgur2,th:godfrey,th:falk}) the
  quantum numbers of the heavy and light quarks are separately
  conserved by the strong interaction.  For the charmed $D$-mesons the
  heavy charmed quark spin, $\mathbf{s_Q}$, couples with the light quark
  momentum \(\mathbf{j_q = s_q + L}\), where $\mathbf{s_q}$ is the spin of the
  light quark and $\mathbf{L}$ is its angular momentum. Hence for $P$-wave
  ($L=1$) mesons we obtain two $j_q = 3/2^{+}$ states, the
  \(J^{P}=2^{+}{\rm ,}\,1^{+}\) states, and two \(j_q = 1/2^{+}\) states, 
  the \(J^{P}=0^{+}{\rm ,}\,1^{+}\) states.  In the
  Heavy Quark Symmetry limit, conservation of parity and $\mathbf{j_q}$ 
  requires that the strong decays 
  \(\Dorbj{(j_q = 3/2^{+})}\to{D^{(*)}(j_q = 1/2^{-})}\,\pi{(J^{P}=0^{-})}\) 
  proceed via $D$-wave while
  \(\Dorbj{(j_q = 1/2^{+})}\to{D^{(*)}(j_q = 1/2^{-})}\,\pi{(J^{P}=0^{-})}\)
  proceed via $S$-wave. Therefore \(\Dorbtwo{(J^{P}=2^{+},j_q = 3/2^{+})}\) 
  decays both to \(\Dstar{(J^{P}=1^{-})}\pi{(J^{P}=0^{-})}\) and to 
  \(D{(J^{P}=0^{-})}\pi{(J^{P}=0^{-})}\) in $D$-wave, in contrast to 
  \(\Dorbone{(J^{P}=1^{+},j_q = 3/2^{+})}\) which decays only into 
  \(\Dstar{(J^{P}=1^{-})}\pi{(J^{P}=0^{-})}\) and again via $D$-wave.
  \(\Dorbprone{(J^{P}=1^{+},j_q = 1/2^{+})}\) converts only into 
  \(\Dstar{(J^{P}=1^{-})}\) emitting a pseudoscalar $\pi$ in $S$-wave, 
  while \( \Dorbzero{(J^{P}=0^{+},j_q = 1/2^{+})}\) decays only to 
  \(D{(J^{P}=0^{-})}\pi{(J^{P}=0^{-})}\) and only in $S$-wave. 
  States decaying via $S$-wave are expected to be broad while those 
  decaying via $D$-wave are expected to be narrow.
\par
  The first observation of excited charmed mesons was made by
  the ARGUS Collaboration (see
  References~\cite{arg:newm,arg:2459,arg:2420,arg:parsons,arg:rhein}) and
  immediately confirmed by CLEO~\cite{cleo1.5:89} and the Tagged Photon
  Spectrometer~(FNAL)~\cite{tps} Collaborations.  The states have
  also been studied at LEP~\cite{delphi,delphi1,aleph}. A high statistics
  analysis has been done with the CLEO-II
  detector~\cite{cleo2:d0,cleo2:dplus}.  The most recent measurements
  by the CLEO-II~\cite{cleo2:d1}, BELLE~\cite{belle:03}, and
  FOCUS~\cite{e687,focus:04} experiments have improved the previous
  ones and extracted the contributions of broad \Dorbprone and
  \Dorbzero states.
\par
  Recently the CDF Collaboration has analyzed the signals of orbital
  \Dorbj-meson states using 210~\invpb of data taken with
  the CDF II Detector. We present here the 
  measurements of the $^{3}P$ orbitally excited neutral \Dorbtwoz and \Dorbonez
  charmed mesons.
\section{Analysis and Tracking Calibration.}
  Our analysis is based on a data sample collected by a trigger
  which selected events with at least two displaced tracks of opposite
  charge, each having an impact parameter measured by the CDF
  silicon detector to be larger than 100\mum and a momentum above
  2.0\gevc. The total momentum of the track pair was required to be
  larger than 5.5\gevc.  This ``impact parameter two track trigger''
  sample is enriched by events with heavy quarks decaying via hadronic
  modes. For example the reconstructed signal of the \Dstarp contains 
  $\sim0.5\times10^6$ events in the peak.
\par
  The $^{3}P$-wave neutral charmed mesons \Dorbjz were
  analyzed in their decay modes to $\Dstarp\pim$ and to
  $\Dp\pim$\footnote[1]{Unless otherwise stated all references
  to the specific charge combination imply the charge conjugate
  combination as well.}.
\par 
  Both \Dorbonez and \Dorbtwoz contribute to the mode 
  $\Dstarp\pim$. Here a \Dstarp was reconstructed from the final state of
  \(\Dz\pip_{soft},\,{\rm with}\,\Dz\to\Km\pip\,{\rm combined~with~a}\,\pip_{soft}\)
  soft track (\(\pt\,>400\mevc\)) where events were triggered 
  by the two hardest tracks of four, and the trigger information 
  was matched with tracks offline. 
  The four-track combination was subjected to the two-dimensional vertex fit
  without mass constraints and the resulting decay length \lxy was required
  to be longer than $500\mum$. 
  The candidates for \Dz
  were required to have mass within $\pm{3\sigma}$ of the $M(\Dz)$ signal. 
  Next, the candidates for \Dstarp were extracted from the mass difference
  $\delta{M}$ spectrum of \(M({\Dz}\pip_{soft})-M({\Dz})\) and again
  from within $\pm{3\sigma}$ of $\delta{M}$ around the narrow 
  (\( \sigma(\delta{M})\sim0.6\mevcc \)) \Dstarp peak.
\par 
  The mode \( \Dorbjz\to\Dp\pim ,\,\Dp\to\Km\pip\pip \) has only
  one narrow \Dorbtwoz that contributes directly. As the \Dp signal has a higher
  combinatorial background, harder cuts were applied. The fitted 
  secondary decay vertex was required to be separated from
  the primary one by applying a $\lxy>1000\mum$ cut. The candidates for \Dp
  were selected from the $\pm{3\sigma}$ window around the $M(\Dp)$ peak. To
  further suppress the background, a cut on momentum 
  \( \pt({\pim} )>800\mevc \) was applied as well.
\par 
  Finally the signals of \Dorbonez and \Dorbtwoz were extracted from
  mass difference spectra
  \[ \Delta{M}=M(({\Dstarp\,or\,\Dp})\pim )\,-\,M({\Dstarp\,or\,\Dp}). \] 
\par
\begin{figure}[h]
\begin{minipage}{12pc}
\includegraphics[width=12pc]{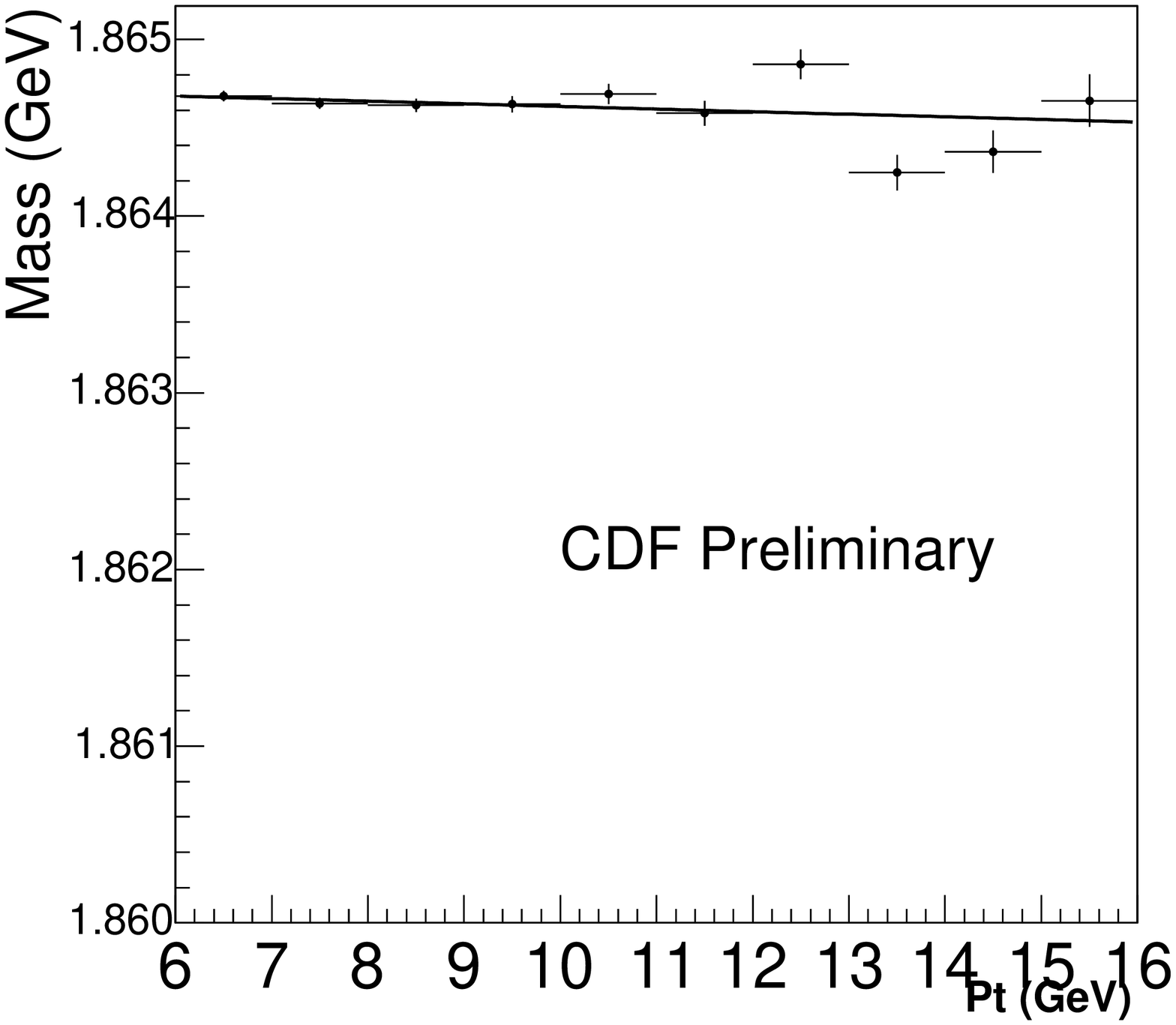}
\caption{\label{cal:d0} 
         \( M(\Dz)_{meas} \) variation versus its momentum \pt,
         \( M(\Dz)_{meas}\in{M(\Dz)_{PDG}}\pm{1\sigma_{PDG}}. \)
        }
\end{minipage}\hspace{0.7pc} 
\begin{minipage}{12pc}
\includegraphics[width=12pc]{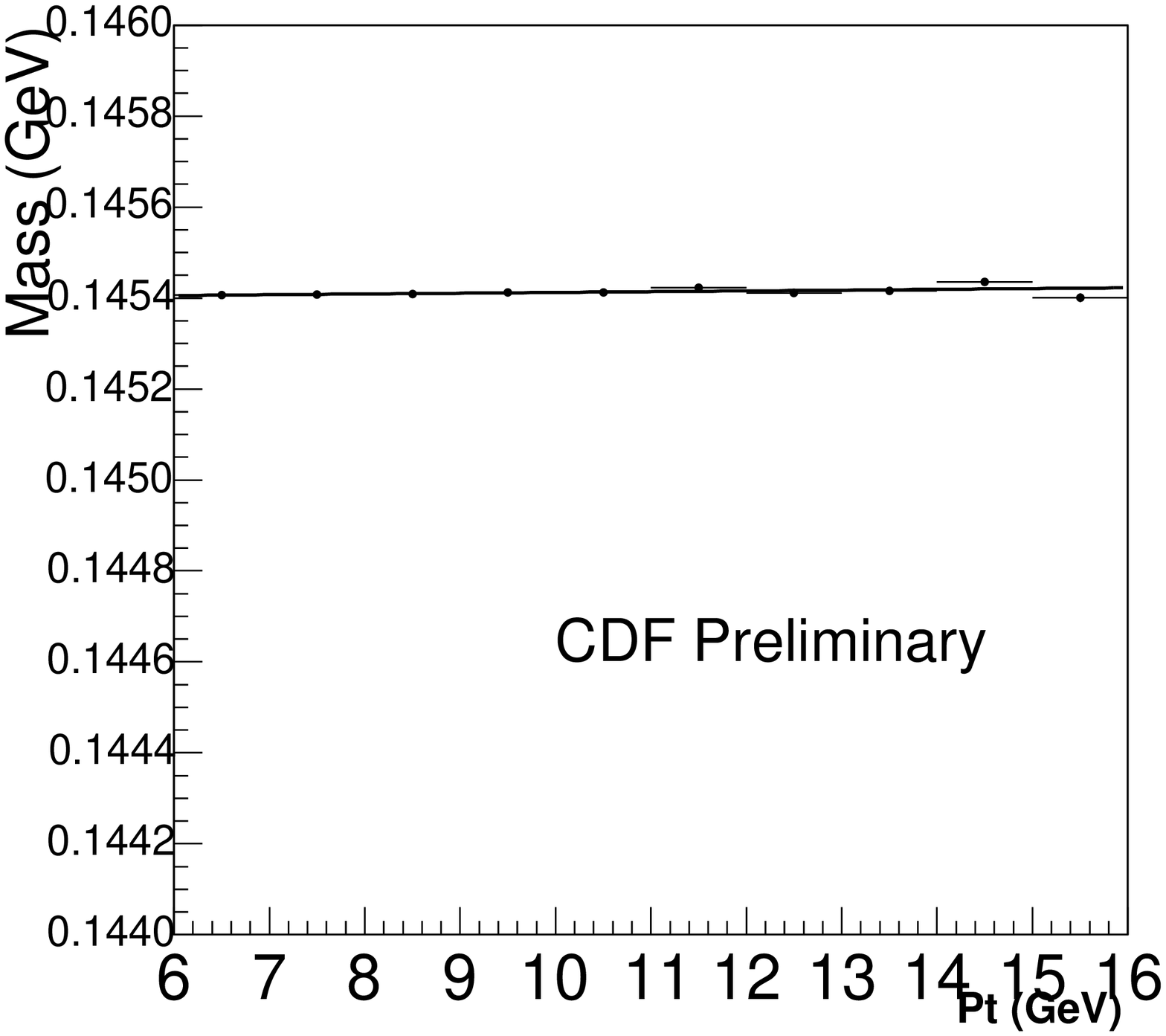}
\caption{\label{cal:ds} 
         \( \delta{M}(\Dstarp)_{meas} \) variation versus its momentum \pt,
         \( \delta{M}(\Dstarp)_{meas}\in{\delta{M}(\Dstarp)_{PDG}}\pm{1\sigma_{PDG}}. \)
        }
\end{minipage}\hspace{0.7pc}
\begin{minipage}{12pc}
\includegraphics[width=12pc]{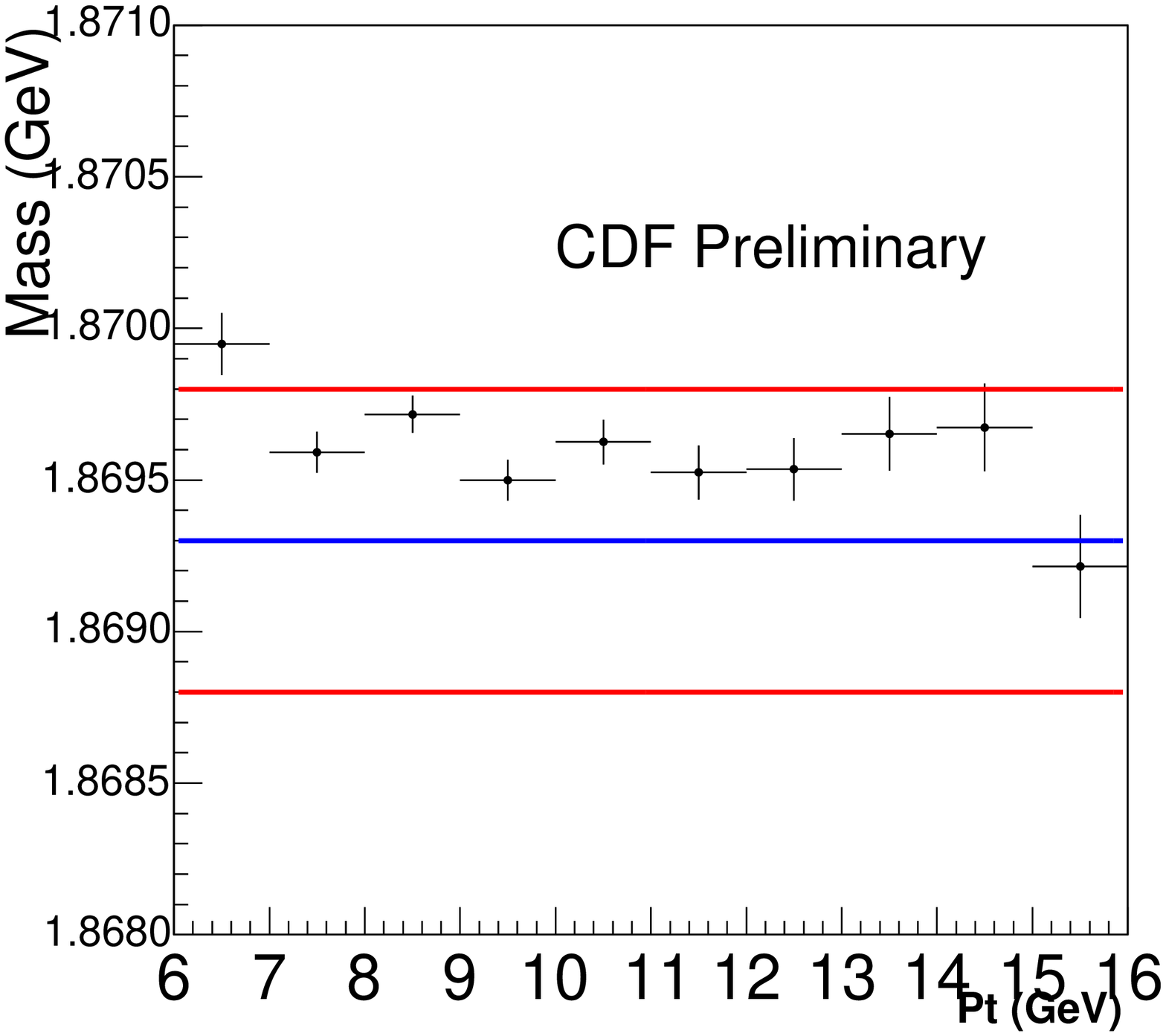}
\caption{\label{cal:dp}
         \( M(\Dp)_{meas} \) variation versus its momentum \pt,
         \( M(\Dp)_{meas}\in{M(\Dp)_{PDG}}\pm{1\sigma_{PDG}}. \)
        }
\end{minipage} 
\end{figure}

 
  A precise measurement of mass requires good 
  calibration of the tracking system. The mass scale was calibrated
  using large samples of $\jpsi\to\mumu$ and $\KS\to\pipi$ events as 
  reference signals.
  The quality of the calibration was validated by checking the variations 
  in mass of the final states, \Dz (see \figref{cal:d0}), \Dstarp (see
  \figref{cal:ds}) and \Dp (see \figref{cal:dp}) against their
  transverse momentum. The variations were found to lie within
  $\pm{1\sigma}$ of the corresponding world averaged
  measurements assigned by the Particle Data Group (PDG)~\cite{pdg2004}.
%
%
\section{Mass Spectra and Fits}
\par
\begin{figure}[h]
\begin{minipage}[b]{18pc}
\includegraphics[width=18pc]{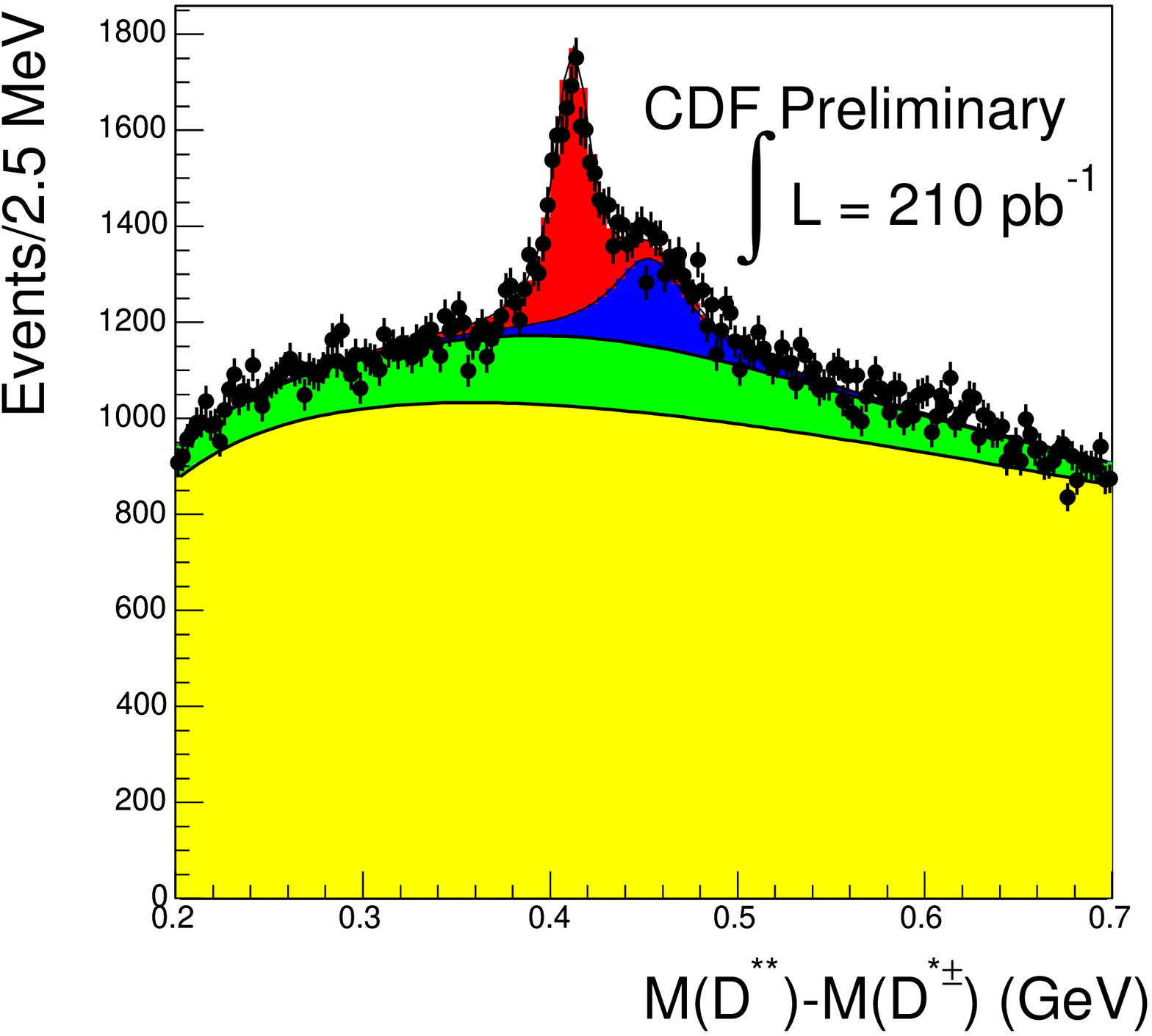}
\caption{\label{sp:dspi}
          Two peaks corresponding to narrow resonances:
          \( \Delta{M(\Dorbonez(2420))}\sim{410\mevcc}\),
          \( \Delta{M(\Dorbtwoz(2460))}\sim{450\mevcc}. \)
        }
\vspace{3.10pc} 
\end{minipage}
\hspace{2pc}%
\begin{minipage}[b]{18pc}
\includegraphics[width=18pc]{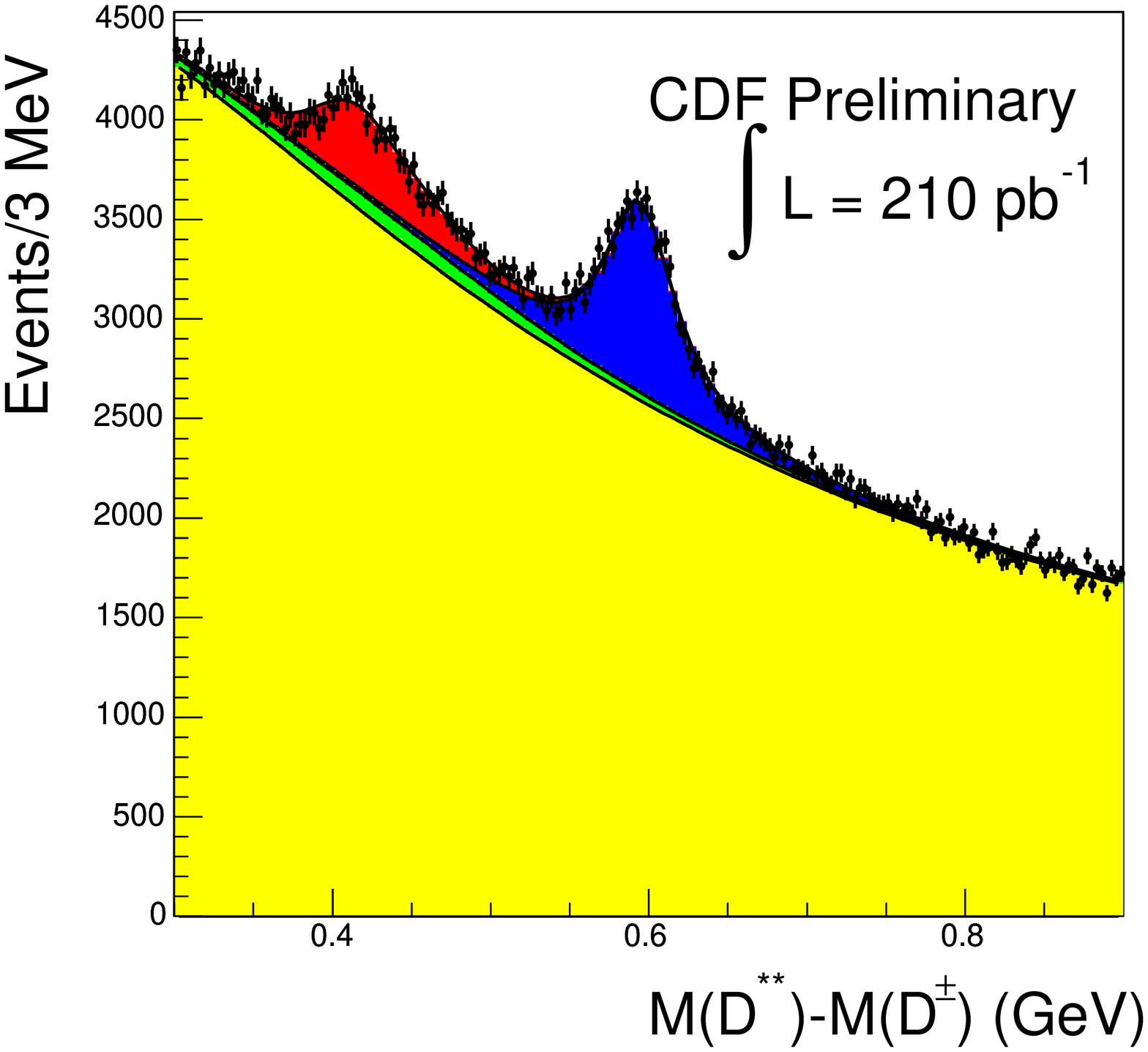}
\caption{\label{sp:dppi}
         Two visible bumps can be attributed to:
         \( \Delta{M(\Dorbtwoz(2460))}\sim{600}\mevcc \)
         and structure at \( \Delta{M}\sim{400}\mevcc \)
         created by the feed-down from \Dstarp\pim states
         when \( \Dstarp\to\Dp\piz(\g)\,\rm{with}\,BR\sim{32}\% \)
         and $\piz(\g)$ goes undetected.
        }
\end{minipage}
\end{figure}

  \figref{sp:dspi} shows the distribution of 
  \( \Delta{M}\equiv\,M({\Dstarp}\pim )\,-\,M({\Dstarp}) \).  A
  double peaked structure is seen with a peak at
  $\sim{410}$\mevcc attributed to the $\Dorbonez(2420)$ and another 
  at $\sim{450}$\mevcc corresponding to the $\Dorbtwoz(2460)$.
\par
  \figref{sp:dppi} shows the mass difference
  spectrum of \( \Delta{M}\equiv\,M({\Dp}\pim )\,-\,M({\Dp}) \).
  The pronounced peak at $\Delta{M}\sim{600}$\mevcc is a signal for
  the $\Dorbtwoz(2460)$ state while the structure below at
  $\Delta{M}\sim{400}$\mevcc is created by feed-down from the 
  $\Dorbonez(2420)$ and $\Dorbtwoz(2460)$ states
  decaying via the \Dstarp\pim mode, with \( \Dstarp\to\Dp\piz(\g) \)
  with a branching ratio of $\sim{32.3}\%$~\cite{pdg2004}.
\par
  The signals are observed on top of a large combinatorial
  background. The broad states contributions, \Dorbpronez in
  \figref{sp:dspi} and \Dorbzeroz in \figref{sp:dppi} are shown as
  well.
\par
  Both $\Delta{M}$ histograms were fitted simultaneously. The
  likelihood function used has independent background and broad state
  \( \Delta{M}(\Dorbpronez\,or\,\Dorbzeroz) \) terms, while the same
  values for masses \( \Delta{M}\) and widths \( \Gamma \) of 
  \Dorbonez\,or\,\Dorbtwoz were kept in the fit.
  The broad and narrow resonances as well as feed-down enhancements
  are described
  by a non-relativistic Breit-Wigner form convoluted with a Gaussian
  resolution function $\mathcal{BW}\otimes{Gauss}$. The resolutions
  $\sigma$ for Gaussian functions were obtained from extensive
  Monte-Carlo studies and fixed in the fit.  The broad states masses
  and widths were fixed to the PDG ones for \Dorbpronez~\cite{pdg2004} or
  to the most recent measurements for \Dorbzeroz~\cite{belle:03}.  The
  fit was tested with a Monte-Carlo sample a factor of 2 larger in size
  than the data, and no selection biases were observed. The corresponding
  statistical uncertainty on fits of Monte-Carlo samples contributes
  to the total systematic error.
%
\section{Results}
\begin{table}[h]
\caption{\label{res:numbers}\cdf2 measurements of \Dorbjz-meson masses and widths,
          shown in {\bf bold style}, compared with other experiments' measurements.}
\begin{center}
\begin{tabular}{lllll}
\hline
{\bf Group}  & {\bf State} & {\bf Mode} & {\bf Mass,\mevcc} & \( {\mathbf \Gamma,\mevcc} \) \\
\hline
CLEO~\cite{cleo2:d0} & \Dorbtwoz & \Dp\pim & 2465$\pm$3$\pm$3       & 28$^{+8}_{-7}\pm\,$6 \\
BELLE~\cite{belle:03}& \Dorbtwoz & \Dp\pim & 2461.6$\pm$2.1$\pm$3.3 & 45.6$\pm$4.4$\pm$6.7 \\
FOCUS~\cite{focus:04}& \Dorbtwoz & \Dp\pim & 2464.5$\pm$1.1$\pm$1.9 & 38.7$\pm$5.3$\pm$2.9 \\
PDG~\cite{pdg2004}   & \Dorbtwoz & \Dstarp\pim, & 2458.9$\pm$2.0    & 23$\pm\,$5 \\
                     &           & \Dp\pim      &                   & \\
\hline
{\bf\cdf2,}&\(\mathbf{\Dorbtwoz}\) &\(\mathbf{\Dp\pim}\) & {\bf 2463.3$\pm$0.6$\pm$0.8} 
                                                         & {\bf 49.2$\pm$2.1$\pm$1.2}\\
                                                         \cline{4-4}
{\bf this }&                       &                     & \(\mathbf{M(\Dorbtwoz)-M(\Dp)}\) & \\
                                                         \cline{4-4}
{\bf study}&                       &                     & {\bf 594.0$\pm$0.6$\pm$0.5} & \\
\hline
CLEO~\cite{cleo2:d0} & \Dorbonez & \Dstarp\pim & 2421$^{+1}_{-2}\pm\,$2 & 20$^{+6}_{-5}\pm$3 \\
BELLE~\cite{belle:03}& \Dorbonez & \Dstarp\pim & 2421.4$\pm$1.5$\pm$0.9 & 23.7$\pm$2.7$\pm$4.0 \\
PDG~\cite{pdg2004}   & \Dorbonez & \Dstarp\pim & 2422.2$\pm\,$1.8       & 18.9$^{+4.6}_{-3.5}$ \\
\hline
{\bf\cdf2,} & \(\mathbf{\Dorbonez}\) & \(\mathbf{\Dstarp\pim}\) & {\bf 2421.7$\pm$0.7$\pm$0.6} 
                                                                & {\bf 20.0$\pm$1.7$\pm$1.3} \\
                                                                \cline{4-4}
{\bf this } &                        &                          & \(\mathbf{M(\Dorbonez)-M(\Dstarp)}\) & \\
                                                                \cline{4-4}
{\bf study} &                        &                          & {\bf 411.7$\pm$0.7$\pm$0.4} & \\
\hline
\end{tabular}
\end{center}
\end{table}
  The fits of our spectra yielded the numbers shown in \tabref{res:numbers}. 
  Results by other high-statistics experiments are shown as well. 
  Our mass difference measurements are also presented.
\par
  The systematic uncertainty on the mass and width measurements
  comprises the uncertainty due to the size of the Monte-Carlo sample used
  to determine the systematic error on the fits, the uncertainty of the masses
  and the natural widths of broad state contributions which were fixed
  in the fits, the uncertainty on the \cdf2 tracker calibration and the
  systematic uncertainty on the magnetic field in the \cdf2 detector. The error
  assigned by the PDG to the masses of the \Dp and \Dstarp does contribute to
  the uncertainty on the absolute mass measurements.
\par
  The \cdf2 mass measurements of the \Dorbtwoz and \Dorbonez charmed
  mesons are in good agreement with world data.  While the width
  $\Gamma(\Dorbonez)$ is in good agreement with world data,
  $\Gamma(\Dorbtwoz)$ lies within 1$\sigma$ of the most
  recent results from BELLE~\cite{belle:03} and FOCUS~\cite{focus:04},
  but does differ from the PDG~2004~\cite{pdg2004} average value.
%
\section{Conclusions}
  The CDF Collaboration presents here the first mass and width
  measurements of neutral $^{3}P$-wave \Dorbonez and \Dorbtwoz charmed
  mesons produced at the Tevatron hadron collider.  Thanks to the
  excellent \cdf2 tracking, the measurements have the lowest
  statistical and systematical uncertainties among current data. The
  mass measurements are in good agreement with the most recent
  experimental data.
%
%
\section*{Acknowledgments}
  The author is grateful to his colleagues from the CDF $B$-Physics
  Working Group, especially to Dr.~E.~Gerchtein for useful suggestions
  and comments made during preparation of this talk. The author would
  like to thank Prof.~Sally~C.~Seidel for support of this work,
  fruitful discussions, and comments. The author thanks 
  Dr.~M.~Hoeferkamp for useful comments.
%
%

\smallskip
%
\end{document}